\documentclass[Physsubmission, Phys]{SciPost}

\hypersetup{
    colorlinks,
    linkcolor={red!50!black},
    citecolor={blue!50!black},
    urlcolor={blue!80!black}
}

\usepackage[bitstream-charter]{mathdesign}
\DeclareSymbolFont{usualmathcal}{OMS}{cmsy}{m}{n}
\DeclareSymbolFontAlphabet{\mathcal}{usualmathcal}

\newcommand{\df}{\mathrm{d}}

\newcommand{\braces}[1]{[\hspace{-0.8mm}[#1]\hspace{-0.8mm}]}

\begin{document}

\begin{center}{\Large \textbf{
Next-to-leading power SCET in Higgs amplitudes induced by light quarks\\
}}\end{center}

\begin{center}
Xing Wang\footnote{The speaker, based on works with Ze Long Liu, Bianka Mecaj, Matthias Neubert and Marvin Schnubel.}
\end{center}

\begin{center}
PRISMA$^+$\! Cluster of Excellence {\rm \&} Mainz Institute for Theoretical Physics, Johannes Gutenberg University, 55099 Mainz, Germany
\\
x.wang@uni-mainz.de
\end{center}

\begin{center}
\today
\end{center}

\definecolor{palegray}{gray}{0.95}
\begin{center}
\colorbox{palegray}{
  \begin{tabular}{rr}
  \begin{minipage}{0.1\textwidth}
    \includegraphics[width=35mm]{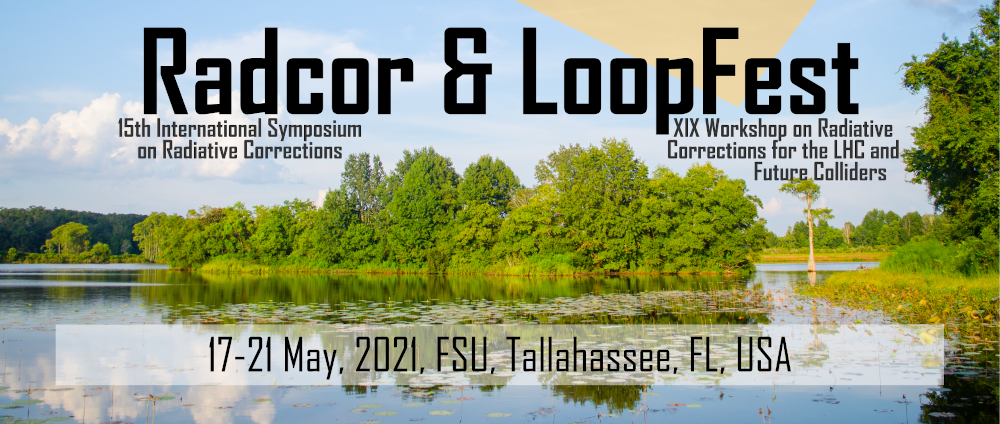}
  \end{minipage}
  &
  \begin{minipage}{0.85\textwidth}
    \begin{center}
    {\it 15th International Symposium on Radiative Corrections: \\Applications of Quantum Field Theory to Phenomenology,}\\
    {\it FSU, Tallahasse, FL, USA, 17-21 May 2021} \\
    \doi{10.21468/SciPostPhysProc.?}\\
    \end{center}
  \end{minipage}
\end{tabular}
}
\end{center}

\section*{Abstract}
{\bf

In this article, we report how to use the ``plus-type" subtraction scheme to deal with endpoint divergences in $h\rightarrow\gamma\gamma$ or $gg\rightarrow h$ amplitudes induced by light quark loop, which can be formulated by next-to-leading power (NLP) SCET. This subtraction is ensured by two re-factorization conditions, which have been proven to all orders in $\alpha_s$. Based on these conditions, cutoffs emerge naturally after some re-arrangements to handle endpoint divergences and renormalization is compatible with that. Our formalism can analytically reproduce three-loop amplitudes up to $\ln^3(-M_h^2/m_b^2-i0)$ and resum to next-to-leading logarithm and beyond. 
}

\vspace{10pt}
\noindent\rule{\textwidth}{1pt}
\tableofcontents\thispagestyle{fancy}
\noindent\rule{\textwidth}{1pt}
\vspace{10pt}

\section{Introduction}
\label{sec:intro}
Soft-collinear effective theory (SCET) \cite{Bauer:2001yt,Bauer:2002nz,Beneke:2002ph} has been proven a powerful tool at leading power to deal with multi-scale problems at colliders. Recently, there have been a lot of works for different processes on understanding the structure beyond the leading order in power counting, including event shape \cite{Moult:2016fqy,Moult:2019vou}, Drell-Yan \cite{Bonocore:2015esa,Beneke:2018gvs,Beneke:2019oqx}, Higgs decay via light quark loops \cite{Liu:2019oav,Wang:2019mym} and off-diagonal splitting function \cite{Beneke:2020ibj}. It turns out that endpoint divergence occurring for convolutions of factorized quantities is a common feature at next-to-leading power (NLP). It makes renormalization and therefore resummation rather non-trivial and is not yet fully understood how to handle it. Recently, Liu and Neubert \cite{Liu:2019oav} found that rapidity divergences in Higgs amplitudes induced by a light quark loop can help to find two re-factorization conditions. We have proven these conditions to all orders in the language of SCET in \cite{Liu:2020ydl}. We have also shown that based on these conditions, endpoint divergences can be cured from the start by a ``plus-type'' subtraction scheme and this scheme is compatible with renormalization. This formalism has been validated to three loops, which is a key step to understand NLP SCET further. We can resum large logarithms to next-to-next-to-leading logarithm (NLL) and even beyond, with the logarithm being the ratio of the Higgs mass and the light quark mass: $L=\ln(-M_h^2/m_b^2-i0)$.  

In this article, we first briefly review the factorization formula for the Higgs amplitude induced by a $b$ quark loop in Sec.~\ref{sec:intro}. Then we present the logic of the subtraction method in Sec.~\ref{sec:subtraction}. Sec.~\ref{sec:RGEandresum} is devoted to illustrate how to renormalize and discuss the RGEs and resummation. We conclude in Sec.~\ref{sec:con}.

\section{Factorization}
In this section, we use $h\rightarrow\gamma\gamma$ by a $b$ quark loop as an example to summarize the operator basis and review the factorization formula at the bare level \cite{Liu:2019oav}. The operator basis and the factorization formula for $gg\rightarrow h$ are rather similar \cite{Liu:2021long}. 

The bare decay amplitude $\mathcal{M}$ contains three terms consisting of unrenormalized SCET operators multiplied (convoluted) by bare Wilson coefficients. The third operator $O_3$ can be further factorized into the convolution of two radiative jet functions and the soft quark soft function. To this end, the bare factorization reads:
\begin{equation}
	\label{eq:barefact}
	\begin{aligned}
		\mathcal{M}=H_1\langle O_1 \rangle + 2\int_0^1\df z\,H_2(z)\langle O_2(z) \rangle + H_3\int_0^\infty\frac{\df\ell_-}{\ell_-}\int_0^\infty\frac{\df\ell_+}{\ell_+}J(M_h\ell_-)J(-M_h\ell_+)S(\ell_-\ell_+).
	\end{aligned}
\end{equation} 
We will call the three terms $T_1^{\text{unsub}},\,T_2^{\text{unsub}}$ and $T_3^{\text{unsub}}$ respectively. The convolutions in the last two terms are a common feature at NLP. On the one hand, unlike the leading power cases where there is only one collinear field at one collinear sector, there can be several collinear fields at one collinear sector. As a result, one has to integrate over the momentum fraction $z$ distributed among these collinear particles\footnote{It is symmetric under $z\leftrightarrow \bar{z}=1-z$.}. On the other hand, there can be some power suppressed interactions now. Hence one has to integrate over the inserting position of such interactions, which is $(\ell_-,\ell_+)$ in this example after multiple expansions in momentum space. 

It turns out that such convolutions are usually divergent when approaching to endpoints: $z\rightarrow 0,\,1$ and $\ell_-,\ell_+\rightarrow \infty$ in \eqref{eq:barefact}. This can be easily shown by noticing that the leading order $\langle O_2(z)\rangle$ is a constant, while the leading order $H_2(z)$ goes like $1/z(1-z)$. Only part of the divergences can be regularized by dimensional regulator, while the remaining ones are rapidity divergences.  Consequently, one can't renormalize the operators as usual and naively replace the bare ones in \eqref{eq:barefact} with the their renormalized counterparts because rapidity divergences still exist when $z\rightarrow 0,\,1$ and $\ell_-,\ell_+\rightarrow \infty$. These superficially troublesome divergences turn to be useful to construct our prescription to cure endpoint divergences.

\section{Plus-type subtraction}
\label{sec:subtraction}
Subtraction is a widely used method to deal with divergences. In this section, we show how to use a ``plus-type'' subtraction scheme to cure endpoint divergences and how it works with renormalization.
\subsection{Re-factorization conditions}
\label{subsec:refact}
As explained in the previous section, both $T_2^{\text{unsub}}$ and $T_3^{\text{unsub}}$ have rapidity divergences. However the amplitude is free of rapidity divergences, which means that there is a cancellation between $T_2^{\text{unsub}}$ and $T_3^{\text{unsub}}$. This indicates that there are some close relations between them.   

\begin{figure}[t]
\begin{center}
\includegraphics[width=0.35\textwidth]{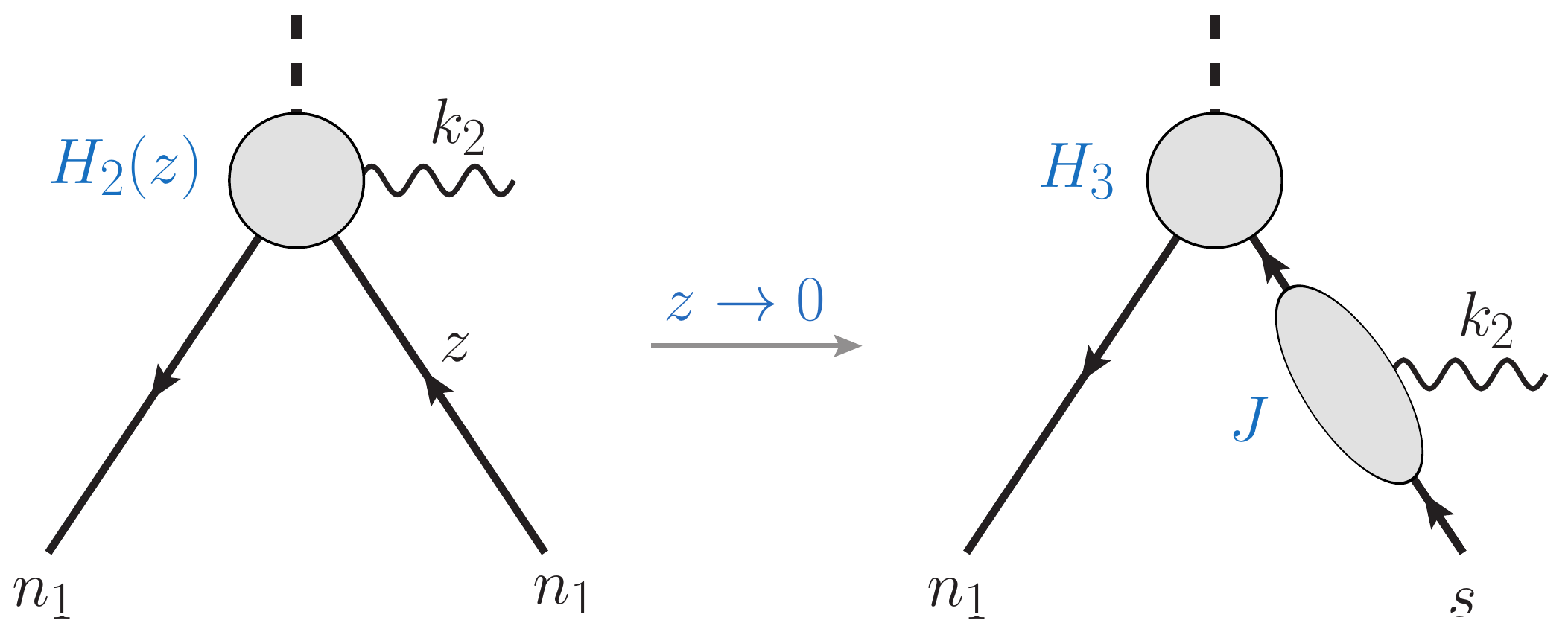}\,\,\,\,\,\,
\includegraphics[width=0.35\textwidth]{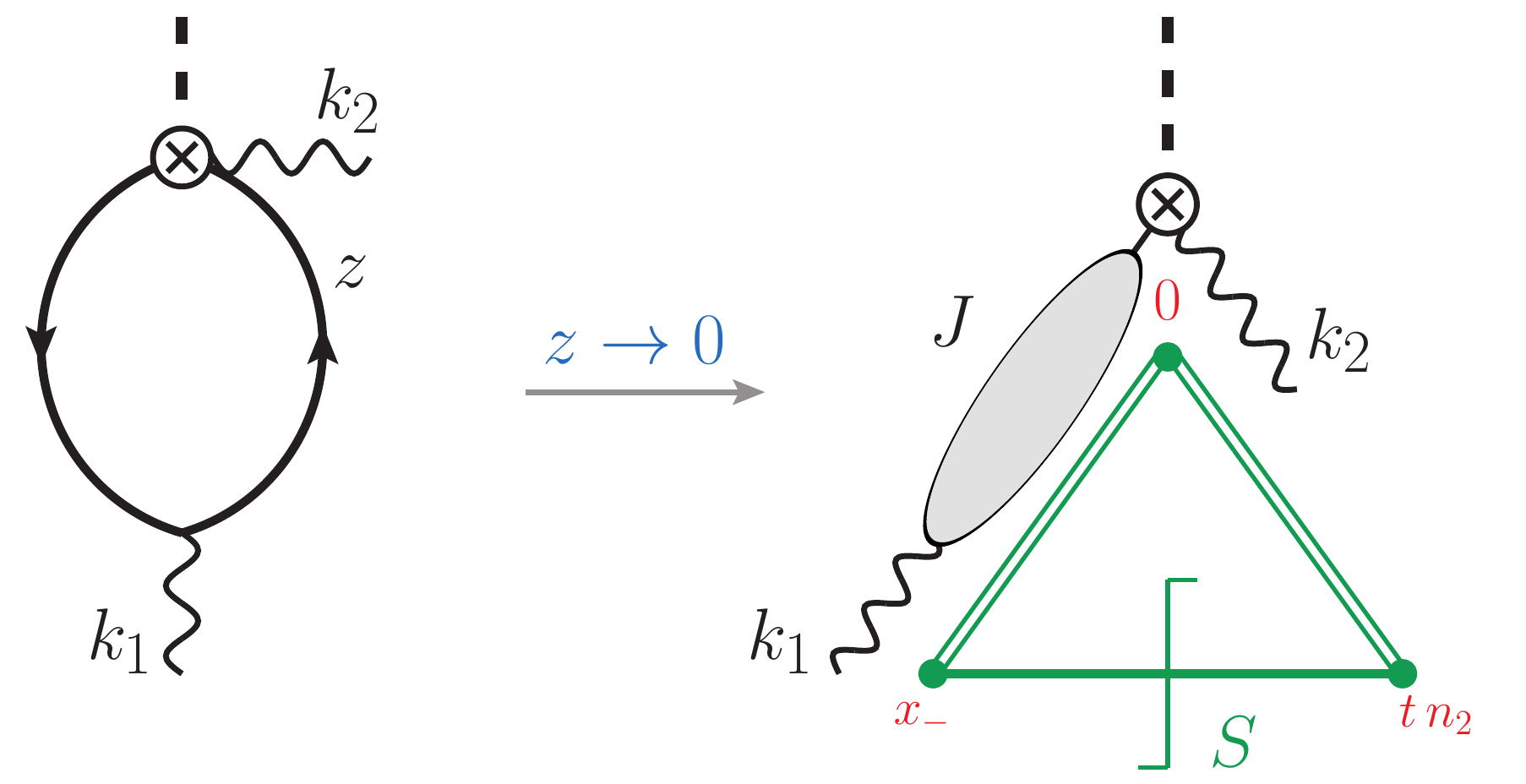} 
\caption{\label{fig:refact} 
Graphical illustration of the two refactorization condition.}
\end{center}
\end{figure}

By using a suitable rapidity regulator \cite{Liu:2019oav}, it has been shown how the rapidity divergences (regularized as $1/\eta$, where $\eta$ is the regulator) cancel among the sum of $T_2^{\text{unsub}}$ and $T_3^{\text{unsub}}$. This is highly non-trivial because $H_2$ and $O_2$ depend on the hard scale $M_h^2$, and the collinear scale $m_b^2$ respectively, while $T_3^{\text{unsub}}$ also involves the object depending on the hard-collinear scale $M_hm_b$, which is the jet function. Objects depending on different scales are related indicates that there are some re-factorization conditions to make the scale separation consistently, just resembles the structure of $O_3$. These conditions were observed and calculated at NLO in \cite{Liu:2019oav}. They were proved to all orders in $\alpha_s$ in \cite{Liu:2020wbn} using the language of SCET. Here we just want to illustrate them by Fig.~\ref{fig:refact} and they read: 
\begin{equation}
	\label{eq:refact}
	\begin{aligned}
		\braces{H_2(z)}=-\frac{H_3}{z}J(zM_h^2),\,\,\,\braces{\langle O_2(z)\rangle}=-\frac{\varepsilon_{\perp}^{*}\left(k_{1}\right) \cdot \varepsilon_{\perp}^{*}\left(k_{2}\right)}{2}\int_0^\infty\frac{\df\ell_+}{\ell_+}J(-M_h\ell_+)S(zM_h\ell_+),
	\end{aligned}
\end{equation}
where $\braces{f(z)}$ means only keeping the leading terms for $z\rightarrow 0$. Note that the second one involves a convolution, such that it only makes sense at the bare level, otherwise it will develop some divergence for $\ell_+\rightarrow\infty$. A similar formula as the first one also appears in \cite{Beneke:2020ibj}. 
For $gg\rightarrow h$, two very similar re-fractorization conditions hold as well. 

\subsection{Subtraction and renormalization}
\label{subsec:renormfact}
Note that $\braces{H_2(z)}$ and $\braces{\langle O_2(z)\rangle}$ are the endpoint versions of $H_2(z)$ and $\langle O_2(z)\rangle$ respectively. This means we can subtract $\braces{H_2(z)}\braces{\langle O_2(z)\rangle}$ from the bare integrand of $T_2^{\text{unsub}}$, such that the subtracted $T_2^{\text{unsub}}$ does not have endpoint divergences any more. Then we have to add the subtracted terms back to $T_3^{\text{unsub}}$. Thanks to the two re-fractorization conditions, the subtraction has the same integrand as in $T_3^{\text{unsub}}$, but with a different integral limit, such that the neat effect is the change of integration upper boundaries from infinity to the hard scale! To this end, the amplitude $\mathcal{M}$ in \eqref{eq:barefact} becomes \cite{Liu:2019oav}:
\begin{equation}
	\label{eq:subtraction}
	\begin{aligned}
		(H_1+\Delta H_1)\langle O_1 \rangle + 2\int_0^1\df z\Big[H_2(z)\langle O_2(z) \rangle-\braces{H_2(z)}\braces{\langle O_2(z)\rangle}-\braces{H_2(\bar{z})}\braces{\langle O_2(\bar{z})\rangle} \Big]\\
		 +\lim_{\sigma\rightarrow -1} H_3\int_0^{M_h}\frac{\df\ell_-}{\ell_-}\int_0^{\sigma M_h}\frac{\df\ell_+}{\ell_+}J(M_h\ell_-)J(-M_h\ell_+)S(\ell_-\ell_+)\Big|_{\text{LP}},
	\end{aligned}
\end{equation}
where the analytic continuation and leading power expansion have to be taken in the end. $\Delta H_1$ is the so-called infinity bin contribution, which has been subtracted twice, so we need to add it back to $H_1$. We now denote the three contributions above as $T_1,\,T_2$ and $T_3$. A key point is that $T_i$'s are now free of endpoint divergences and we no longer need rapidity regulator. Note that \eqref{eq:subtraction} is still at the bare level.  

Now turn to renormalization of \eqref{eq:subtraction}. Notice that $O_3$ is a time product between the scalar current and the NLP Lagrangian, which does not need renormalization. So $O_3$ renormalizes like the scalar current via $Z_{33}$, which has been known to three loops \cite{Becher:2009qa}. Then it is straightforward to get $H_3(\mu)=Z_{33}^{-1}H_3$. We know that $O_3$ factorizes as a convolution of the jet functions and the soft function and the jet function also appears in the factorization formula of $B^-\rightarrow \gamma \ell^{-} \bar{\nu}$. Consequently $Z_J$ can be deduced from the RG consistency of this decay amplitude. Recently its renormalization has also been studied directly in \cite{Bodwin:2021epw}. Then we deduced $Z_S$ for the soft function from the information of $O_3$ and the jet function and it was confirmed directly in \cite{Bodwin:2021cpx}. Again thanks to the first re-fractorization condition in \eqref{eq:refact}, we also know the $Z-$factor for $\braces{H_2(z)}$. 
\begin{equation}
	\label{eq:ZJZS}
	\begin{aligned}
		J(p^2,\mu)=\int_0^\infty\df x\,Z_J(p^2,xp^2)J(xp^2),\,\,\,S(w,\mu)=\int\limits_0^\infty\df w^\prime Z_S(w,w^\prime)S(w^\prime).
	\end{aligned}
\end{equation}

The $Z-$factor of $O_2(z)$ can be calculated directly, and the corresponding diagrams are given in \cite{Liu:2020wbn}. It turns out that it is not only renormalized multiplicatively, but it mixes with $O_1$ under renormalization as well. Since the colour fields in $O_2$ are the same as those in the leading-twist light-cone distribution amplitude of a transversely polarized vector meson, it is not surprising that the well-known Brodsky-Lepage kernel enters here. By taking the endpoint limit, we can directly obtain the factors for $\braces{O_2(z)}$, whose non-mixing term is nothing but the inverse of that of $\braces{H_2(z)}$, serving as a crosscheck. 
\begin{equation}
	\label{eq:O2}
	\begin{aligned}
		\langle O_2(z,\mu)\rangle=&\int_0^1\df z^\prime\, Z_{22}(z,z^\prime)\langle O_2(z^\prime)\rangle+Z_{21}(z)\langle O_1\rangle ,\,\\
		\braces{\langle O_2(z,\mu)\rangle}=&\int_0^\infty\df z^\prime\, \braces{Z_{22}(z,z^\prime)}\braces{\langle O_2(z^\prime)\rangle}+\braces{Z_{21}(z)}\langle O_1\rangle ,\,\\
		H_2(z,\mu)=&\int_0^1\df z^\prime\,H_2(z^\prime)Z_{22}^{-1}(z^\prime,z),\,\\
		\braces{H_2(z,\mu)}=&\int_0^\infty\df z^\prime\,\braces{H_2(z^\prime)}\braces{Z_{22}^{-1}(z^\prime,z)}.
	\end{aligned}
\end{equation}

Renormalizing $O_1$ is trivial: $\langle O_1(\mu)\rangle=Z_{11}\langle O_1\rangle$, but renormalizing $H_1$ is the most complicated throughout this project. This complexity originates from the fact that there are no cutoffs when renormalizing, see \eqref{eq:ZJZS}, but there exist cutoffs in the convolutions of the factorization formula \eqref{eq:subtraction} both for $T_3$ and the subtracted terms in $T_2$. As a result, when we re-express the bare quantities in $T_2$ and $T_3$ in terms of their renormalized counterparts, there will be some extra contribution, because exchanging the integration boundaries does not commute here (or there is mismatch for the integral boundaries) We call them ``left-over'' terms. By a careful proof to all orders in $\alpha_s$, we find that the sum of ``left-over'' terms in $T_2$ and $T_3$, denoted as $\delta H_1^{\text{tot}}$, is purely hard, which can be naturally absorbed into $H_1$. This happens because of the two re-fractorization conditions. On top of that, the renormalized $H_1$ also obtains contribution from operator mixing in \eqref{eq:O2}. To this end, we arrive at the formula:
\begin{equation}
	\label{eq:H1}
	\begin{aligned}
		H_1(\mu)=\Big[H_1+\Delta H_1+\delta H_1^{\text{tot}}\Big]Z_{11}^{-1}+2\int\limits_0^1\df z\Big(H_2(z)Z_{21}^{-1}(z)-\braces{H_2(z)}\braces{Z_{21}^{-1}(z)}\\
		-\braces{H_2(\bar{z})}\braces{Z_{21}^{-1}(\bar{z})} \Big),
	\end{aligned}
\end{equation}
where $Z_{21}^{-1}$ is calculated from $Z_{22}^{-1}$, $Z_{21}$ and $Z_{11}^{-1}$, likewise for $\braces{Z_{21}^{-1}}$.

At the end of the day, we get the factorization formula with renormalized quantities:
\begin{equation}
	\label{eq:renorfact}
	\begin{aligned}
		T_1^R=&H_1(\mu)\langle O_1(\mu)\rangle, \,\\
		T_2^R=&2\int_0^1\df z\Big[H_2(z,\mu)\langle O_2(z,\mu) \rangle-\braces{H_2(z,\mu)}\braces{\langle O_2(z,\mu)\rangle}-\braces{H_2(\bar{z},\mu)}\braces{\langle O_2(\bar{z},\mu)\rangle} \Big],\,\\
		 T_3^R=&\lim_{\sigma\rightarrow -1} H_3(\mu)\int_0^{M_h}\frac{\df\ell_-}{\ell_-}\int_0^{\sigma M_h}\frac{\df\ell_+}{\ell_+}J(M_h\ell_-,\mu)J(-M_h\ell_+,\mu)S(\ell_-\ell_+,\mu)\Big|_{\text{LP}}.
	\end{aligned}
\end{equation}
Now the amplitude is just $T_1^R+T_2^R+T_3^R$, and it is free of any divergences. At first sight, the cutoffs may seem artificial. However they are not. They emerge as a result of the subtracted terms. Subtracted terms in $T_2$ and $T_3$ are the same up to a minus sigh, which is the result of re-factorization conditions, which are of physical nature of this problem. Emergence of cutoffs come at a price when writing down the renormalized factorization formula, but again re-factorization conditions play a key role.   

\section{Renormalization and resummation}
\label{sec:RGEandresum}
It is straightforward to derive the anomalous dimensions from the renormalization factors\cite{Becher:2005pd}. And based on the knowledge of two-loop anomalous dimension of the jet function, we can go even beyond one-loop except $O_2(z,\mu)$. We present results for $S(w,\mu)$ and $H_1(\mu)$ as two examples. For the soft function, the RGE reads
\begin{equation}
	\label{eq:RGESoft}
	\begin{aligned}
		\frac{\df\,S(w,\mu)}{\df\ln\mu}=-\int_0^\infty\df x\,\gamma_S(w,w/x)S(w/x,\mu),
	\end{aligned}
\end{equation}
with\footnote{Here the cusp anomalous dimension is in the fundamental representation.}
\begin{equation}
	\label{eq:ADSoft}
	\begin{aligned}
		\gamma_S(w,w/x)=-\left[\Gamma_{\text {cusp }}\left(\alpha_{s}\right) \ln \frac{w}{\mu^{2}}-\gamma_{s}\left(\alpha_{s}\right)\right] \delta(1-x)-2 \Gamma_{\text {cusp }}\left(\alpha_{s}\right) \Gamma(1, x)-2 \left(\frac{\alpha_{s}}{4 \pi}\right)^{2} h(x),
	\end{aligned}
\end{equation}
up to $\mathcal{O}(\alpha_s^3)$, where $\Gamma(1,x)$ is the Lange-Neubert kernel \cite{Lange:2003ff} and $h(x)$ has recently been derived in \cite{Braun:2019wyx}. At one-loop\footnote{In this report, we always adopt the convention: $\gamma_V(\alpha_s)=\sum_{i\geq 0}\big(\alpha_s/(4\pi)\big)^{i+1}\gamma_{V,i}$.} $\gamma_{s,0}=-6C_F$. One can find the explicit expression for $\gamma_s$ at two-loop in \cite{Liu:2020eqe,Liu:2020wbn}. The solution of this equation has been studied in \cite{Liu:2020eqe} at RG improved NLO accuracy. An interesting mathematical structure appears due to the two non-local kernels. After resummation, the soft function becomes continuous at the threshold point $w=m_b^2$. It is worthwhile to compare with the anomalous dimension of the soft quark soft function for $gg\rightarrow h$. The soft function for
$gg\rightarrow h$ differs with that for $h\rightarrow\gamma\gamma$ by insertion of two colour matrices, which changes the UV behaviour. The RGE takes the same structure as \eqref{eq:RGESoft} but with a different anomalous dimension. At one-loop order, it reads
\begin{equation}
	\label{eq:ADSoftgg}
	\begin{aligned}
		\gamma^{gg}_S(w,w/x)=-\frac{\alpha_s}{\pi}\Bigg\{\left[(C_F-C_A)\ln\frac{w}{\mu^2}+\frac{3C_F-\beta_0}{2}\right]\delta(1-x)+2\left(C_F-\frac{C_A}{2}\right)\Gamma(1,x)\Bigg\}.
	\end{aligned}
\end{equation} 
We define $r_\Gamma$ as the ratio between the coefficient of the logarithmic term and that of the Lange-Neubert kernel. For $h\rightarrow\gamma\gamma$, $r_\Gamma=2$, it allows the RG solution to be expressed in terms of Meijer-G function. But for $gg\rightarrow h$, $r_\Gamma=1/5$, the solution is not even single-valued in the complex plane. The same story also goes for the jet function. But for NLL resummation, we can simplify the result by Taylor expansion. 
To give a taste of this comparison for resummed soft functions at RG improved LO, we go to the regime $w\gg m_b^2$, where the solutions simplify significantly:
\begin{equation}
	\label{eq:Softres}
	\begin{aligned}
		S_{\text{RGi}}^{\text{LO}}(w,\mu)=&-N_c\frac{\alpha_b}{\pi}m_b(\mu_s)\left(\frac{w}{\mu_s^2}\right)^{-a_\Gamma}e^{2S_\Gamma+a_{\gamma_s}}\left(\frac{\Gamma(1+a_{\Gamma})}{\Gamma(1-a_{\Gamma})}e^{2\gamma_E a_\Gamma}\right)^2,\,\\
		S_{\text{RGi}}^{gg,\text{LO}}(w,\mu)=&-T_F\delta_{ab}\frac{\alpha_s(\mu_s)}{\pi} m_b(\mu_s)\left(\frac{w}{\mu_s^2}\right)^{-a_{\Delta\Gamma}}e^{2S_{\Delta\Gamma}+a_{\gamma^{\prime}_s}}\left(\frac{\Gamma(1+a_{\Delta\Gamma})}{\Gamma(1-a_{\Delta\Gamma})}e^{2\gamma_E a_{\Delta\Gamma}}\right)^{2r_\Gamma},
	\end{aligned}
\end{equation}
where $\gamma_{s,0}=-6C_F,\,\gamma^\prime_{s,0}=-(6C_F-2\beta_0)$ are used to compute the RG function $a_V$. The RG functions above are defined as:
\begin{equation}
    \label{eq:RGfunctions}
	\begin{aligned}
	S_{V}(\nu,\mu)=-\int\limits_{\alpha_s(\nu)}^{\alpha_s(\mu)}\df\alpha\,\frac{\gamma_V(\alpha)}{\beta(\alpha)}\int\limits_{\alpha_s(\nu)}^{\alpha}\frac{\df\alpha^\prime}{\beta(\alpha^\prime)},\,\,\, a_{ V}(\nu,\mu)=-\int\limits_{\alpha_s(\nu)}^{\alpha_s(\mu)}\,\df\alpha\frac{ \gamma_V(\alpha)}{\beta(\alpha)}\,,
	\end{aligned}
\end{equation}
where $V$ can be any anomalous dimensions above, for example, $a_{\Delta\Gamma}$ takes $\Delta\Gamma=\alpha_s(C_F-C_A)/\pi+\cdots$ and $S_{\Gamma}$ takes $\Gamma=\alpha_sC_F/\pi+\cdots$. And to save space, the arguments in the RG functions are suppressed. The full solution can be found in \cite{Liu:2020eqe,Liu:2021jet,Liu:2021long}

The RGE of $H_1(\mu)$ is again complicated due to its involved definition \eqref{eq:H1}. With the knowledge of all the renormalization factors, at last, we find that 
\begin{equation}
	\label{eq:RGEH1}
	\begin{aligned}
		\frac{\df H_1(\mu)}{\df\ln\mu}=\gamma_{11}H_1(\mu)+D_{\text{cut}}(\mu)+2\int_0^1\df z\Big(H_2(z,\mu)\gamma_{21}(z)-\braces{H_2(z,\mu)}\braces{\gamma_{21}(z)}\\
		-\braces{H_2(\bar{z},\mu)}\braces{\gamma_{21}(\bar{z})}\Big),
	\end{aligned}
\end{equation}
where the last term is obviously the mixing contribution, while the second term comes from the ``left-over'' contribution. We don't give the definition for $D_{\text{cut}}(\mu)$ here. But we want to emphasize that $D_{\text{cut}}(\mu)\ni \alpha_s^2 L_h^2 $, so the RGE of $H_1(\mu)$ is not Sudakov-like. To solve this kind of RGE is still an open question, but we don't have to solve it to resum large logarithms, because we can choose to evolve the operators up to the hard scale $\mu_h$. A slightly different RGE holds for the case of $gg\rightarrow h$ \cite{Liu:2021long}. 

The solutions to the RGEs of $\langle O_2(z,\mu)\rangle $ and $\braces{\langle O_2(z,\mu)\rangle}$ are more involved due to the subtraction nature. We have found a way to solve them systematically by a semi-numerical method. They are not needed if we care about next-to-leading logarithmic (NLL) resummation of the amplitude. At this accuracy, all contributions come from $T_3$, as a subset of RG improved leading order $T_3$. To be more concrete, at NLL, the asymptotic solution \eqref{eq:Softres} is enough for the soft function. Combined with the resummed jet and hard function, and Taylor expanding the integrand in $T_3$, we arrive at:
\begin{equation}
	\label{eq:NLLintegrand}
	\begin{aligned}
		T_3^{\text{NLL}}=N_c\frac{\alpha_b}{\pi}\frac{y_b(\mu_h)}{\sqrt{2}}m_b(\mu_h)\int\limits_0^\infty\df L_-\int\limits_0^\infty\df L_+\theta(L-L_--L_+)\exp\Bigg\{-\frac{\alpha_s(\mu_h)}{4\pi}\times\\
		 \Bigg[\frac{\Gamma_0}{4}\bigg(L_s^2-L_-^2-L_+^2+\frac{\alpha_s(\mu_h)}{4\pi}\frac{\beta_0}{3}(L_s^3-L_-^3-L_+^3)\bigg)+\frac{\gamma_{s,0}+\gamma_{m,0}}{2}L_s\Bigg]\Bigg\},
	\end{aligned}
\end{equation}   
where $\gamma_{m,0}=-6C_F,\,\Gamma_0=4C_F,\,\beta_0=11/3C_A-4/3T_Fn_f$ and $L_s=L_-+L_+$. After the integration, we obtain (9) in \cite{Liu:2020tzd}. For $gg\rightarrow h$, a similar formula can be found in \cite{Liu:2021long}.

We can also use the RGEs to predict large logarithms in the three-loop amplitudes, from $L^6$ to $L^3$. After switching to on-shell scheme, and making the choice $\mu^2=\mu_h^2=-M_h^2-i0$, we find that the amplitude takes the form:
\begin{equation}
	\label{eq:amp}
	\begin{aligned}
\mathcal{M}=& \frac{N_{c} \alpha_{b}}{\pi} \frac{m_{b}y_b}{\sqrt{2}} \varepsilon_{\perp}^{*}\left(k_{1}\right) \cdot \varepsilon_{\perp}^{*}\left(k_{2}\right)\Bigg\{\frac{L^{2}}{2}-2 \\
 &+\frac{C_{F} \alpha_{s}\left(\mu_{h}\right)}{4 \pi}\left[-\frac{L^{4}}{12}-L^{3}-\frac{2 \pi^{2}}{3} L^{2}+\left(12+\frac{2 \pi^{2}}{3}+16 \zeta_{3}\right) L-20+4 \zeta_{3}-\frac{\pi^{4}}{5}\right]\\
&+C_{F}\left(\frac{\alpha_{s}\left(\mu_{h}\right)}{4 \pi}\right)^{2}\left[\frac{C_{F}}{90} L^{6}+\left(\frac{C_{F}}{10}-\frac{\beta_{0}}{30}\right) L^{5}+d_{4}^{\mathrm{OS}} L^{4}+d_{3}^{\mathrm{OS}} L^{3}+\ldots\right]\Bigg\},
\end{aligned}
\end{equation} 
where
\begin{equation}
	\label{eq:ampcont}
	\begin{aligned}
	d_{4}^{\mathrm{OS}}=&\left(\frac{3}{2}+\frac{\pi^{2}}{18}\right) C_{F}+\left(-\frac{91}{27}+\frac{\pi^{2}}{36}\right) C_{A}+\frac{32}{27} T_{F} n_{f}, \\
	d_{3}^{\mathrm{OS}}=&\left(-\frac{1}{2}+\frac{7 \pi^{2}}{9}+\frac{20}{3} \zeta_{3}\right) C_{F}+\left(-\frac{199}{18}-\frac{22 \pi^{2}}{27}-4 \zeta_{3}\right) C_{A}+\left(\frac{22}{9}+\frac{8 \pi^{2}}{27}\right) T_{F} n_{f}.
\end{aligned}
\end{equation}
This is in perfect agreement with \cite{Niggetiedt:2020sbf} after decoupling the coupling constant. We have done the same thing for $gg\rightarrow h$ case, even though we don't know two loop anomalous dimension for the jet and the soft function exactly \cite{Liu:2021long}. We have found perfect agreement with \cite{Spira:1995rr,Aglietti:2006tp} at NLO and \cite{Czakon:2020vql} at NNLO, which is a strong cross-check of our formalism and the RGEs.

\section{Conclusion}
\label{sec:con}
In this proceeding, we report our recent progress on NLP SCET, in the context of Higgs amplitudes induced by a light quark loop (both for colourless and colourful external states). SCET can be used to derive the factorization formulae systematically. We find that the ``plus-type'' subtraction scheme to cure endpoint divergences naturally is compatible with renormalization. We derive the RGEs of factorized quantities and obtain the solutions. New functional forms appear after resummation and new colour structures appear surprisingly. Based on the RGEs, we can resum large logarithms to NLL, and beyond to RG improved LO. We also reproduce three-loop amplitudes analytically up to $\mathcal{O}(L^3)$, in agreement with QCD three-loop numeric calculations.


\paragraph{Funding information}
This material is based upon works partially supported by the Cluster of Excellence PRISMA$^+$ funded by the German Research Foundation (DFG) within the German Excellence Strategy (Project ID 39083149).


\begin{thebibliography}{10}
\providecommand{\url}[1]{\texttt{#1}}
\providecommand{\urlprefix}{URL }
\expandafter\ifx\csname urlstyle\endcsname\relax
  \providecommand{\doi}[1]{doi:\discretionary{}{}{}#1}\else
  \providecommand{\doi}{doi:\discretionary{}{}{}\begingroup
  \urlstyle{rm}\Url}\fi
\providecommand{\eprint}[2][]{\url{#2}}

\bibitem{Bauer:2001yt}
C.~W. Bauer, D.~Pirjol and I.~W. Stewart,
\newblock \emph{{Soft collinear factorization in effective field theory}},
\newblock Phys. Rev. D \textbf{65}, 054022 (2002),
\newblock \doi{10.1103/PhysRevD.65.054022},
\newblock \eprint{hep-ph/0109045}.

\bibitem{Bauer:2002nz}
C.~W. Bauer, S.~Fleming, D.~Pirjol, I.~Z. Rothstein and I.~W. Stewart,
\newblock \emph{{Hard scattering factorization from effective field theory}},
\newblock Phys. Rev. D \textbf{66}, 014017 (2002),
\newblock \doi{10.1103/PhysRevD.66.014017},
\newblock \eprint{hep-ph/0202088}.

\bibitem{Beneke:2002ph}
M.~Beneke, A.~P. Chapovsky, M.~Diehl and T.~Feldmann,
\newblock \emph{{Soft collinear effective theory and heavy to light currents
  beyond leading power}},
\newblock Nucl. Phys. B \textbf{643}, 431 (2002),
\newblock \doi{10.1016/S0550-3213(02)00687-9},
\newblock \eprint{hep-ph/0206152}.

\bibitem{Moult:2016fqy}
I.~Moult, L.~Rothen, I.~W. Stewart, F.~J. Tackmann and H.~X. Zhu,
\newblock \emph{{Subleading Power Corrections for N-Jettiness Subtractions}},
\newblock Phys. Rev. D \textbf{95}(7), 074023 (2017),
\newblock \doi{10.1103/PhysRevD.95.074023},
\newblock \eprint{1612.00450}.

\bibitem{Moult:2019vou}
I.~Moult, G.~Vita and K.~Yan,
\newblock \emph{{Subleading power resummation of rapidity logarithms: the
  energy-energy correlator in $ \mathcal{N} $ = 4 SYM}},
\newblock JHEP \textbf{07}, 005 (2020),
\newblock \doi{10.1007/JHEP07(2020)005},
\newblock \eprint{1912.02188}.

\bibitem{Bonocore:2015esa}
D.~Bonocore, E.~Laenen, L.~Magnea, S.~Melville, L.~Vernazza and C.~D. White,
\newblock \emph{{A factorization approach to next-to-leading-power threshold
  logarithms}},
\newblock JHEP \textbf{06}, 008 (2015),
\newblock \doi{10.1007/JHEP06(2015)008},
\newblock \eprint{1503.05156}.

\bibitem{Beneke:2018gvs}
M.~Beneke, A.~Broggio, M.~Garny, S.~Jaskiewicz, R.~Szafron, L.~Vernazza and
  J.~Wang,
\newblock \emph{{Leading-logarithmic threshold resummation of the Drell-Yan
  process at next-to-leading power}},
\newblock JHEP \textbf{03}, 043 (2019),
\newblock \doi{10.1007/JHEP03(2019)043},
\newblock \eprint{1809.10631}.

\bibitem{Beneke:2019oqx}
M.~Beneke, A.~Broggio, S.~Jaskiewicz and L.~Vernazza,
\newblock \emph{{Threshold factorization of the Drell-Yan process at
  next-to-leading power}},
\newblock JHEP \textbf{07}, 078 (2020),
\newblock \doi{10.1007/JHEP07(2020)078},
\newblock \eprint{1912.01585}.

\bibitem{Liu:2019oav}
Z.~L. Liu and M.~Neubert,
\newblock \emph{{Factorization at subleading power and endpoint-divergent
  convolutions in $h\to\gamma\gamma$ decay}},
\newblock JHEP \textbf{04}, 033 (2020),
\newblock \doi{10.1007/JHEP04(2020)033},
\newblock \eprint{1912.08818}.

\bibitem{Wang:2019mym}
J.~Wang,
\newblock \emph{{Resummation of double logarithms in loop-induced processes
  with effective field theory}}  (2019),
\newblock \eprint{1912.09920}.

\bibitem{Beneke:2020ibj}
M.~Beneke, M.~Garny, S.~Jaskiewicz, R.~Szafron, L.~Vernazza and J.~Wang,
\newblock \emph{{Large-x resummation of off-diagonal deep-inelastic parton
  scattering from d-dimensional refactorization}},
\newblock JHEP \textbf{10}, 196 (2020),
\newblock \doi{10.1007/JHEP10(2020)196},
\newblock \eprint{2008.04943}.

\bibitem{Liu:2020ydl}
Z.~L. Liu and M.~Neubert,
\newblock \emph{{Two-Loop Radiative Jet Function for Exclusive $B$-Meson and
  Higgs Decays}},
\newblock JHEP \textbf{06}, 060 (2020),
\newblock \doi{10.1007/JHEP06(2020)060},
\newblock \eprint{2003.03393}.

\bibitem{Liu:2021long}
Z.~L. Liu, M.~Neubert, S.~Marvin and X.~Wang,
\newblock \emph{{Factorization, Renormalization and Endpoint Divergences at
  Subleading Power in $gg\to h$ Production}},
\newblock to be appear .

\bibitem{Liu:2020wbn}
Z.~L. Liu, B.~Mecaj, M.~Neubert and X.~Wang,
\newblock \emph{{Factorization at subleading power and endpoint divergences in
  $h\to\gamma\gamma$ decay. Part II. Renormalization and scale evolution}},
\newblock JHEP \textbf{01}, 077 (2021),
\newblock \doi{10.1007/JHEP01(2021)077},
\newblock \eprint{2009.06779}.

\bibitem{Becher:2009qa}
T.~Becher and M.~Neubert,
\newblock \emph{{On the Structure of Infrared Singularities of Gauge-Theory
  Amplitudes}},
\newblock JHEP \textbf{06}, 081 (2009),
\newblock \doi{10.1088/1126-6708/2009/06/081},
\newblock [Erratum: JHEP 11, 024 (2013)],
\newblock \eprint{0903.1126}.

\bibitem{Bodwin:2021epw}
G.~T. Bodwin, J.-H. Ee, J.~Lee and X.-P. Wang,
\newblock \emph{{Renormalization of the radiative jet function}}  (2021),
\newblock \eprint{2107.07941}.

\bibitem{Bodwin:2021cpx}
G.~T. Bodwin, J.-H. Ee, J.~Lee and X.-P. Wang,
\newblock \emph{{Analyticity, renormalization, and evolution of the soft-quark
  function}},
\newblock Phys. Rev. D \textbf{104}(1), 016010 (2021),
\newblock \doi{10.1103/PhysRevD.104.016010},
\newblock \eprint{2101.04872}.

\bibitem{Becher:2005pd}
T.~Becher and M.~Neubert,
\newblock \emph{{Toward a NNLO calculation of the anti-B ---\ensuremath{>} X(s)
  gamma decay rate with a cut on photon energy: I. Two-loop result for the soft
  function}},
\newblock Phys. Lett. B \textbf{633}, 739 (2006),
\newblock \doi{10.1016/j.physletb.2006.01.006},
\newblock \eprint{hep-ph/0512208}.

\bibitem{Lange:2003ff}
B.~O. Lange and M.~Neubert,
\newblock \emph{{Renormalization group evolution of the B meson light cone
  distribution amplitude}},
\newblock Phys. Rev. Lett. \textbf{91}, 102001 (2003),
\newblock \doi{10.1103/PhysRevLett.91.102001},
\newblock \eprint{hep-ph/0303082}.

\bibitem{Braun:2019wyx}
V.~M. Braun, Y.~Ji and A.~N. Manashov,
\newblock \emph{{Two-loop evolution equation for the B-meson distribution
  amplitude}},
\newblock Phys. Rev. D \textbf{100}(1), 014023 (2019),
\newblock \doi{10.3204/PUBDB-2019-02451},
\newblock \eprint{1905.04498}.

\bibitem{Liu:2020eqe}
Z.~L. Liu, B.~Mecaj, M.~Neubert, X.~Wang and S.~Fleming,
\newblock \emph{{Renormalization and Scale Evolution of the Soft-Quark Soft
  Function}},
\newblock JHEP \textbf{07}, 104 (2020),
\newblock \doi{10.1007/JHEP07(2020)104},
\newblock \eprint{2005.03013}.

\bibitem{Liu:2021jet}
Z.~L. Liu, M.~Neubert, S.~Marvin and X.~Wang,
\newblock \emph{{Two-loop Radiative Gluon Jet Function for $gg\rightarrow h$
  via Light-quark Loop and the Scale Evolution}},
\newblock to be appear .

\bibitem{Liu:2020tzd}
Z.~L. Liu, B.~Mecaj, M.~Neubert and X.~Wang,
\newblock \emph{{Factorization at subleading power, Sudakov resummation, and
  endpoint divergences in soft-collinear effective theory}},
\newblock Phys. Rev. D \textbf{104}(1), 014004 (2021),
\newblock \doi{10.1103/PhysRevD.104.014004},
\newblock \eprint{2009.04456}.

\bibitem{Niggetiedt:2020sbf}
M.~Niggetiedt,
\newblock \emph{{Exact quark-mass dependence of the Higgs-photon form factor at
  three loops in QCD}},
\newblock JHEP \textbf{04}, 196 (2021),
\newblock \doi{10.1007/JHEP04(2021)196},
\newblock \eprint{2009.10556}.

\bibitem{Spira:1995rr}
M.~Spira, A.~Djouadi, D.~Graudenz and P.~M.~Zerwas,
\newblock \emph{{Higgs boson production at the LHC}},
\newblock Nucl. Phys. B \textbf{453} (1995), 17-82,
\newblock \doi{10.1016/0550-3213(95)00379-7},
\newblock \eprint{hep-ph/9504378}.

\bibitem{Aglietti:2006tp}
U.~Aglietti, R.~Bonciani, G.~Degrassi and A.~Vicini,
\newblock \emph{{Analytic Results for Virtual QCD Corrections to Higgs Production and Decay}},
\newblock JHEP \textbf{01} (2007), 021,
\newblock \doi{10.1088/1126-6708/2007/01/021},
\newblock \eprint{hep-ph/0611266}.


\bibitem{Czakon:2020vql}
M.~Niggetiedt,
\newblock \emph{{Exact quark-mass dependence of the Higgs-gluon form factor at three loops in QCD}},
\newblock JHEP \textbf{05} (2020), 149,
\newblock \doi{10.1007/JHEP05(2020)149},
\newblock \eprint{2001.03008}.


\end{thebibliography}

\nolinenumbers

\end{document}